\documentclass[reprint,superscriptaddress,aps,prl]{revtex4-1}
\usepackage{graphicx}
\usepackage{dcolumn}
\usepackage{bm}
\usepackage{bbold}
\usepackage{amsfonts,color}
\usepackage{amsmath}

\newcounter{lastnote}

\begin{document}

\title{Non-Hermitian coherent coupling of nanomagnets by exchange spin waves}

\author{Hanchen Wang}
\thanks{These authors contributed equally to this work.}
\affiliation{%
Fert Beijing Institute, BDBC, School of Microelectronics, Beihang University, Beijing 100191, China
}
\author{Jilei Chen}
\thanks{These authors contributed equally to this work.}
\affiliation{%
Fert Beijing Institute, BDBC, School of Microelectronics, Beihang University, Beijing 100191, China
}%
\author{Tao Yu}
\thanks{These authors contributed equally to this work.}
\affiliation{%
Max Planck Institute for the Structure and Dynamics of Matter, 22761 Hamburg, Germany
}%
\author{Chuanpu Liu}
\thanks{These authors contributed equally to this work.}
\affiliation{%
Fert Beijing Institute, BDBC, School of Microelectronics, Beihang University, Beijing 100191, China
}%
\author{Chenyang Guo}
\thanks{These authors contributed equally to this work.}
\affiliation{%
Beijing National Laboratory for Condensed Matter Physics, Institute of Physics, University of Chinese Academy of Sciences, Chinese Academy of Sciences, Beijing 100190, China
}%
\author{Hao Jia}
\thanks{These authors contributed equally to this work.}
\affiliation{%
Shenzhen Institute for Quantum Science and Engineering (SIQSE), and Department of Physics, Southern University of Science and Technology (SUSTech), Shenzhen 518055, China
}%
\author{Song Liu}
\thanks{These authors contributed equally to this work.}
\affiliation{%
Shenzhen Institute for Quantum Science and Engineering (SIQSE), and Department of Physics, Southern University of Science and Technology (SUSTech), Shenzhen 518055, China
}%
\author{Ka Shen}
\thanks{These authors contributed equally to this work.}
\affiliation{%
Department of Physics, Beijing Normal University, Beijing 100875, China
}%
\author{Tao~Liu}
\affiliation{%
Department of Physics, Colorado State University, Fort Collins, Colorado 80523, USA
}%
\author{Jianyu Zhang}
\affiliation{%
Fert Beijing Institute, BDBC, School of Microelectronics, Beihang University, Beijing 100191, China
}%
\author{Marco A. Cabero Z}
\affiliation{%
Shenzhen Institute for Quantum Science and Engineering (SIQSE), and Department of Physics, Southern University of Science and Technology (SUSTech), Shenzhen 518055, China
}%
\affiliation{%
Fert Beijing Institute, BDBC, School of Microelectronics, Beihang University, Beijing 100191, China
}%
\author{Qiuming Song}
\affiliation{%
Shenzhen Institute for Quantum Science and Engineering (SIQSE), and Department of Physics, Southern University of Science and Technology (SUSTech), Shenzhen 518055, China
}%
\author{Sa Tu}
\affiliation{%
Fert Beijing Institute, BDBC, School of Microelectronics, Beihang University, Beijing 100191, China
}%
\author{Mingzhong~Wu}
\affiliation{%
Department of Physics, Colorado State University, Fort Collins, Colorado 80523, USA
}%
\author{Xiufeng Han}
\affiliation{%
Beijing National Laboratory for Condensed Matter Physics, Institute of Physics, University of Chinese Academy of Sciences, Chinese Academy of Sciences, Beijing 100190, China
}%
\author{Ke Xia}
\affiliation{%
Shenzhen Institute for Quantum Science and Engineering (SIQSE), and Department of Physics, Southern University of Science and Technology (SUSTech), Shenzhen 518055, China
}%
\author{Dapeng Yu}
\affiliation{%
Shenzhen Institute for Quantum Science and Engineering (SIQSE), and Department of Physics, Southern University of Science and Technology (SUSTech), Shenzhen 518055, China
}%
\author{Gerrit E. W. Bauer}
\affiliation{%
Kavli Institute of Nanoscience, Delft University of Technology, 2628 CJ Delft, The Netherlands
}%
\affiliation{Institute for Materials Research, WPI-AIMR and CSNR, Tohoku University, Sendai 980-8577, Japan}%
\affiliation{%
Zernike Institute for Advanced Materials, University of Groningen, Nijenborgh 4, 9747 AG Groningen, The Netherlands
}%
\author{Haiming Yu}
\email{haiming.yu@buaa.edu.cn}
\affiliation{%
Fert Beijing Institute, BDBC, School of Microelectronics, Beihang University, Beijing 100191, China
}


\begin{abstract}
Non-Hermitian physics has recently attracted much attention in optics and photonics. Less explored is non-Hermitian magnonics that provides opportunities to take advantage of the inevitable dissipation of magnons or spin waves in magnetic systems. Here we demonstrate non-Hermitian coherent coupling of two distant nanomagnets by fast spin waves with sub-50 nm wavelengths. Magnons in two nanomagnets are unidirectionally phase-locked with phase shifts controlled by magnon spin torque and spin-wave propagation. Our results are attractive for analog neuromorphic computing that requires unidirectional information transmission. 
\end{abstract}


\maketitle
The conservation of energy is a cornerstone of physics and requires Hermitian Hamiltonians with real eigenvalues~\cite{Morse1953}. However, when a subsystem receives from or loses energy to the environment, its Hamiltonian becomes non-Hermitian~\cite{Christ2018}. Non-Hermitian optics and photonics~\cite{Miri2019} exhibits fascinating phenomena, such as exceptional point singularities~\cite{Heiss2012}, unidirectional invisibility~\cite{Lin2011} and spontaneous parity-time symmetry breaking~\cite{Feng2014}. Spin waves or its quanta, the magnons, are collective excitations of the magnetic order that are studied in the field called magnonics~\cite{Vla2008,Krug2010,Len2011,Vogt2014,Chu2015}. Magnons are strongly coupled with photons in microwave cavities~\cite{Zhang2014,Bai2015} that allows them to hybridize with the charge excitation of a superconducting qubit~\cite{Huebl2013,Tabuchi2015} and reveal macroscopic quantum effects~\cite{Lach2020}. 

Non-Hermitian magnonics is still in its infancy in comparison with its photonic counterpart. Magnons and photons are bosons~\cite{Dem2006,Sch2020} but with different dispersion relations: at the same frequencies magnons have much shorter wavelengths than photons, which gives spin waves an innate advantage for integrated nanodevices~\cite{Khitun2010,Wagner2016,Halder2016,Grundler2016,Csaba2020}. The dynamics of spatially separated nanomagnets can be synchronized by the exchange of magnons, which has important technological implications for e.g. microwave generators~\cite{Kaka2005,Madami2011,Demi2012,Uraz2014,Zahe2020}. Recently, chirally coupled nanomagnets by the Dzyaloshinskii-Moriya interaction has also been reported~\cite{Luo2019}.

Here, we experimentally demonstrate a non-Hermitian coupling of two distant ultra-thin film Co nanowires ($>$50 nm wide) that is mediated by fast exchange spin waves in 20 nm-thick yttrium iron garnet (YIG) films. The excitation and detection of coherent exchange spin waves is non-trivial by itself~\cite{Wintz2016,Hama2017,Liu2018,Die2019,Che2020}. In this work, coherent exchange spin waves are excited with wavelengths down to $\lambda=48$ nm, which are fastest reported so far~\cite{Wintz2016,Hama2017,Liu2018,Die2019,Che2020,SI}. More importantly, the Co wire are coherently and uni-directionally coupled by magnon spin torque~\cite{Slon2010,Han2019,Wang2019,Weiler2018,Chen2018,Qin2018,Li2020} mediated by spin waves with a phase shift $\Delta \phi$ that is freely tunable from 0 to 2$\pi$ by propagation-induced phase delay. The coupled system of two Co wires and magnetic substrate can be formulated in terms of an effective non-Hermitian Hamiltonian~\cite{Chen2019,TYu2019,TYu_arXiv}. The unidirectional dissipative coupling also strongly enhances the spin-wave mediated transmission signal~\cite{Xia2018}, quite different from the level attraction of magnon and photon levels reported in a microwave cavity~\cite{Hu2018}. 

\begin{figure*}[!ht]
\centering
\includegraphics[width=178mm]{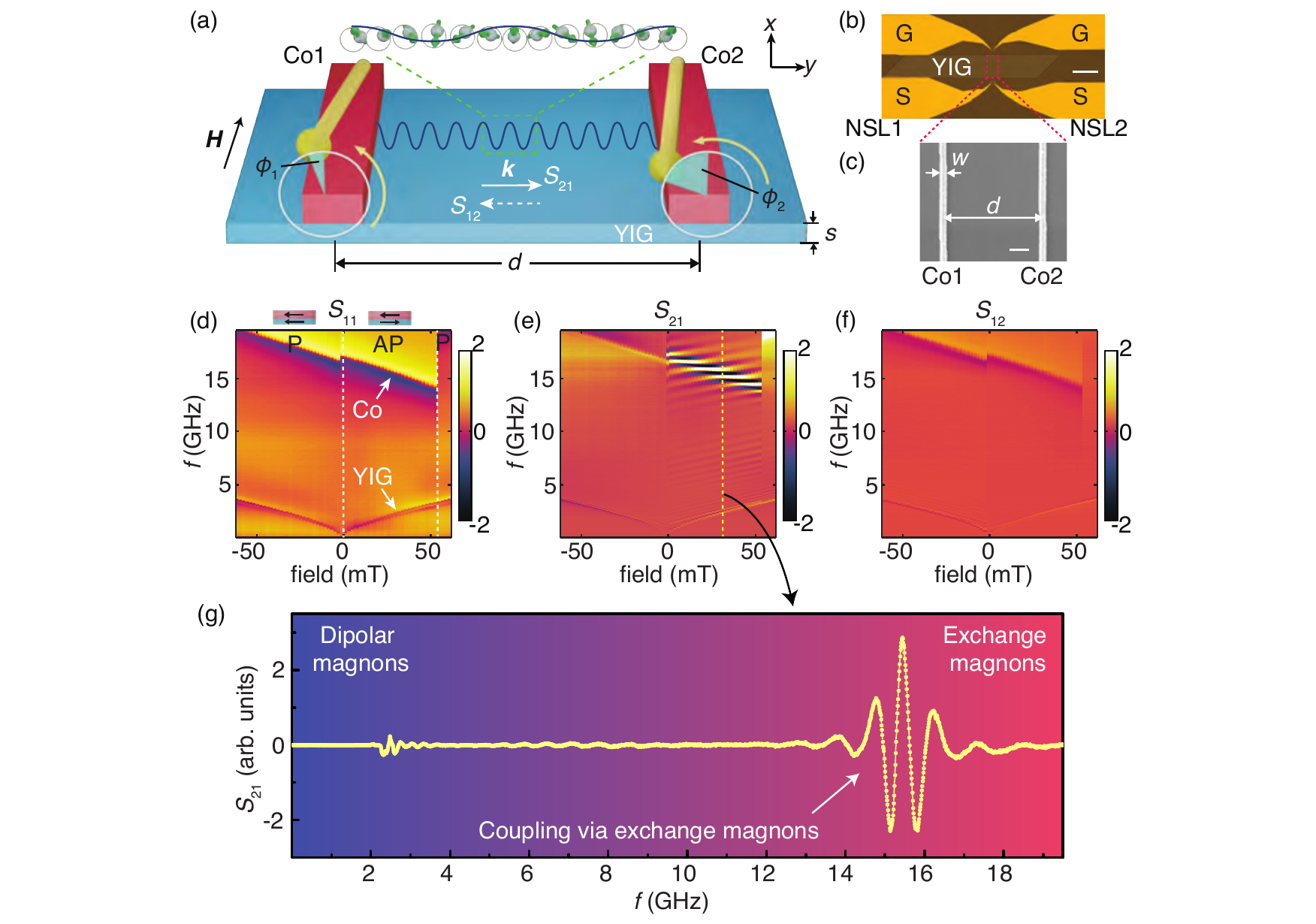}
\caption{(a) A schematic diagram of two Co nanowires coupled by spin waves in the YIG thin film. Parameters are the thickness of the film $s$, the distance between the nanowires d, the spin-wave wavevector k and the in-plane applied field H. The coupling between Co1 (2) and Co2 (1) is evident in the microwave transmission spectra $S_{21}$ ($S_{12}$) measured by the nano-stripline antennas. The shift $\Delta\phi=\phi_{2}-\phi_{1}$ of the precession phases of the Kittel modes in the two wires can be controlled. (b) An optical microscope image of the device with two ground-signal antennas, NSL1 and NSL2. Scale bar, 100~$\mu$m. (c) An SEM image of two Co wires with width $w=100$~nm and center-to-center distance $d=1.5$~$\mu$m. The scale bar indicates 300~nm. (d) The reflection spectrum $S_{11}$ measures the absorption of microwaves generated by NSL1. The ferromagnetic resonances of the Co wire and YIG film are indicated by white arrows. By sweeping the magnetic field H from negative to positive values, the magnetization of the Co/YIG bilayer switches from a parallel (P) to an antiparallel (AP) magnetic configuration in the interval from 0 to 54~mT marked by white dashed lines. The (imaginary component of the) microwave transmission spectra $S_{21}$ plotted in (e) exhibits strong spin-wave oscillations in the AP state close to the Co resonance. In the reversed transmission $S_{12}$ (f) such oscillations are not detected. (g) is a line plot of the spectrum at 30~mT, tracing the yellow dashed line in (e), in which the blue and red areas represent the dipolar and exchange magnon regimes, respectively. The enhanced transmission signal indicated by the white arrow is caused by the coupling of the two Co wires through exchange magnons in the film.}
\label{fig1}
\end{figure*}

Figure~\ref{fig1}(a) illustrates two identical Co nanowires in direct contact with a 20 nm-thin yttrium iron garnet (YIG) film. The Co wires are 30 nm thin, 100 nm wide and separated by 1.5 $\mu$m. A scanning electron microscope (SEM) image is shown in Fig.~\ref{fig1}(c). Magnons in two Co wires are coupled through spin waves in the YIG film with low magnetic damping~\cite{Chang2014,Yu2014}. On each Co wires, gold nano-stripline (NSL) antennas (200 nm wide)~\cite{Ciubotaru2016} are integrated to coherently excite and detect spin waves (Fig.~\ref{fig1}(b)). The microwave reflection spectrum $S_{11}$ (Fig.~\ref{fig1}(d)) is sensitive to the ferromagnetic resonance of the material close to NSL1. The Co wire Kittel mode is observed at frequencies (14.5$\sim$19.5 GHz) much higher than that of the YIG film (0.5$\sim$3.5 GHz) due to the large form anisotropy of Co nanowires~\cite{Topp2010,Ding2011}. An in-plane magnetic field $H$ is applied parallel to the Co wires. By sweeping $H$ from negative to positive values, the magnetic configuration of Co/YIG device switches from parallel (P) to antiparallel (AP) because of the high coercivity of Co nanowires~\cite{Topp2010,Ding2011} compared to the YIG film, returning to a parallel configuration only at H$>$50~mT. 

Microwaves are transmitted from Co1 to Co2 via propagating spin waves~\cite{Vla2008,Neu2010,Yu2014,Han2019} as measured by the transmission spectra $S_{21}$ (its imaginary part is shown in Fig.~\ref{fig1}(e)). We observe remarkable interference fringes (ripples) around the Co resonance in the AP state. The features in $S_{21}$ (Fig.~\ref{fig1}(e)) vanish in the reverse transmission spectra $S_{12}$ (Fig.~\ref{fig1}(f)), revealing a nearly perfect chiral excitation of exchange spin waves by the magnetodipolar fields from the Co nanowires~\cite{Chen2019,TYu2019,TYu_arXiv}. This can be understood by the stray fields generated by the right (left) moving spin waves that vanish below (above) the film and precess in the opposite direction of the magnetization. Figure~\ref{fig1}(g) is a cross section from $S_{21}$ extracted at a field of 30 mT. The Co FMR related signal near 15~GHz is more than 10 times stronger than the signal at lower frequencies assigned to dipolar magnons close the FMR of YIG~\cite{Sun2012,Serga2010}. Therefore, the short-wavelength, exchange-dominated magnons with high group velocities~\cite{Wintz2016,Hama2017,Liu2018,Die2019,Che2020} transmit signals between two Co wires at high frequencies (14$\sim$17 GHz) and in one direction only.

\begin{figure*}[!ht]
\includegraphics[width=178mm]{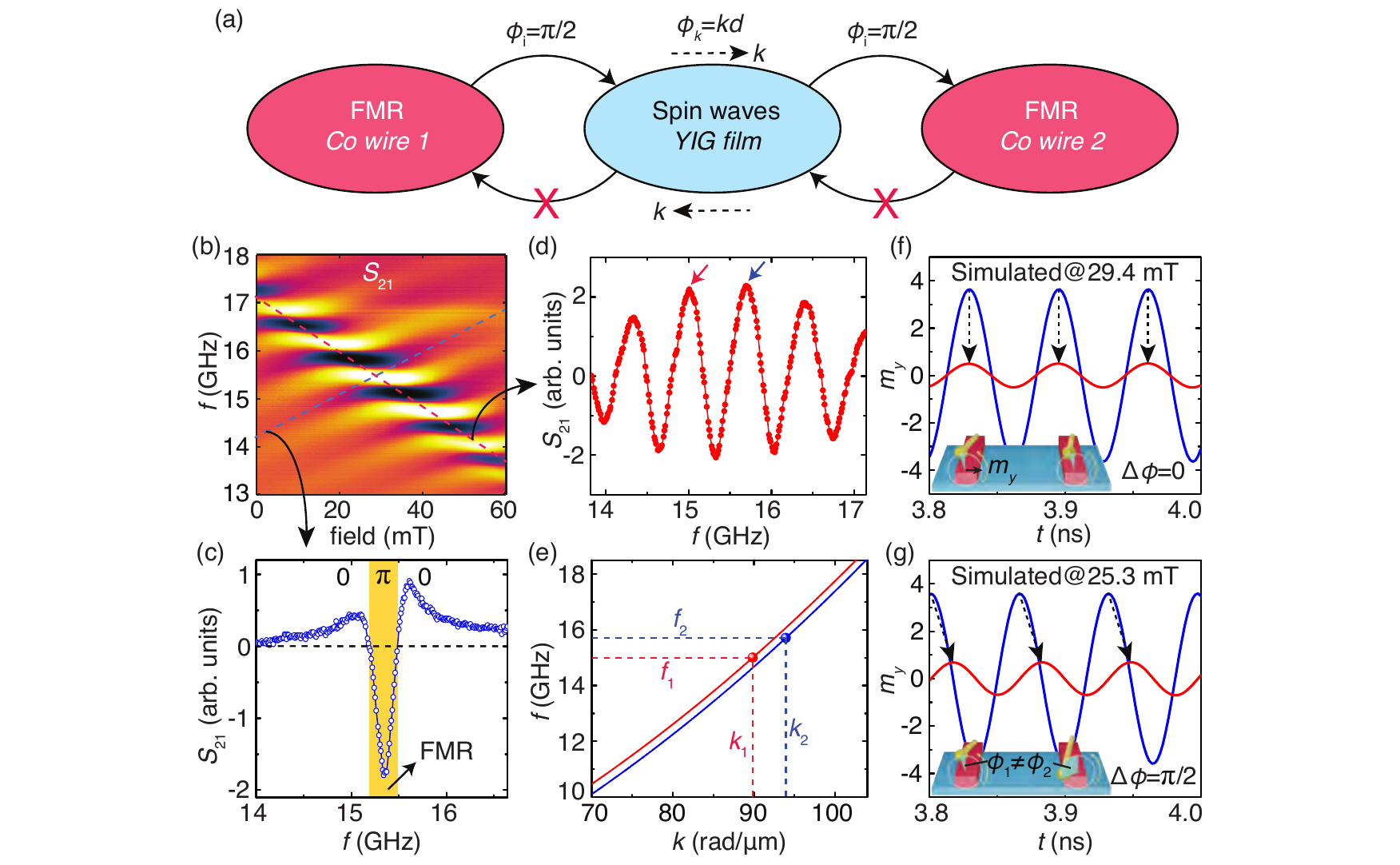}
\caption{(a) A sketch of the phase-locking observed between two Co wires on top of a YIG film. A Co wire under resonance imprints a dissipative phase shift $\phi_{\text{i}}=\pi/2$ on exciting spin waves below by the dipolar interaction. The phase delay due to the propagation of spin waves is given by $\phi_{k}=kd$. This coupling is unidirectional, which implies a non-Hermitian coupling. (b) Transmission spectra $S_{21}$ measured at AP states of Co/YIG (close-up of Fig.~\ref{fig1}(e)). (c) A cut of the spectrum along the blue dashed line in (b) shows a phase change of $\pi$ in the vicinity of the Co resonance (yellow region). (d) is $S_{21}$  along the Co resonance, following the red-dashed line in (b). The frequencies of the transmission maxima $f_1=15.00$~GHz and $f_2=15.71$~GHz at the red and blue arrows correspond to wave numbers $k_1=89.9$~rad/$\mu$m and $k_2=93.9$~rad/$\mu$m using the calculated spin-wave dispersion (e) at magnetic fields of 26.0 mT (blue curve) and 38.5 mT (red curve). (f) The results of micromagnetic simulations for the $y$ component of the magnetization $m_{y}$ of Co1 (blue curve) and Co2 (red curve) as a function of $t$ for 29.5~mT and 15.25~GHz. The black arrows indicated zero phase change, i.e. a perfect synchronization. (g) As (f), but for 25.3~mT and 15.30~GHz, leading to $\Delta\phi=\pi/2$. 
}
\label{fig2}
\end{figure*}

We find a fixed phase relation $\Delta\phi=\phi_{1}-\phi_{2}$ between the spin precessions of the two Co wires that is caused by: (1) Twice a shift of $\phi_{\text{i}}$ by the dynamical coupling~\cite{Weiler2018,Chen2018,Qin2018,Li2020} between the Co wires and the YIG film, i.e. from Co1 to YIG film and from YIG film to Co2, which depends on the frequency. At resonance this process is purely dissipative, i.e. $\phi_{\text{i}}=\pi/2$; (2) the phase delay $\phi_k$ caused by the propagation of exchange spin waves with a finite wavenumber $k$ over a distance $d$. The total phase-difference between the Kittel modes of Co2 and Co1, thus reads~\cite{SI}
\begin{equation}\label{phase_shift}
\begin{split}
	\left\langle\hat{m}_{2}\right\rangle &= \frac{2\left|\sigma_{k}\right|}{\frac{1}{2}\kappa_{\text{Co}}+\left|\sigma_{k}\right|}e^{i\Delta\phi}\left\langle\hat{m}_{1}\right\rangle\\\
	\Delta\phi &= 2\phi_{\text{i}}+\phi_{k}\\
	\phi_{\text{i}} &= \pi/2\\
	\phi_{k} &= kd
\end{split}
\end{equation}
in which $\left\langle\hat{m}_{2}\right\rangle$($\left\langle\hat{m}_{1}\right\rangle$) are the coherent amplitudes of the Kittel magnons in Co2 (Co1), $\sigma_{k}$ is the additional damping induced by the interface Zeeman coupling~\cite{SI}, $\kappa_{\text{Co}}=2\omega\alpha_{\text{G}}$ the reciprocal lifetime proportional to the Gilbert damping parameter $\alpha_{\text{G}}$ of the Co wires. The blue dashed line traces a mode of exchange spin waves with short wavelengths propagating in YIG thin film~\cite{Liu2018}. When the propagating spin-wave mode crosses the Co resonance (red dashed line), the phase shifts by $\pi$ and the transmission amplitude changes sign (yellow area of Fig.~\ref{fig2}(c)). The non-locally excited Co2 can again pump spin waves (to the right) with a phase $\Delta\phi=3\phi_{\text{i}}+\phi_{k}$ that destructively interferes with the waves coming from Co1 that have not been absorbed by Co2. Figure~\ref{fig2}(d) shows a spectrum extracted along the Co FMR (red dashed line in~\ref{fig2}(b)) with $2\phi_{\text{i}}=\pi$. The periodicity of the oscillations is caused by the propagation phase delay $\phi_{k}=kd$. Between two neighboring peaks (marked by red and blue arrows) $\Delta\phi=2\pi$. Since the propagation distance d is fixed, the phase change is governed by the wavenumber variation $\Delta k=k_{2}-k_{1}$ for the peak frequencies $f_{1}$ and $f_{2}$. The dispersion relation of exchange-dipolar spin waves~\cite{Kal1986} in the Damon-Eshbach configuration for the parameters of a 20 nm-thick YIG film~\cite{Chang2014,Yu2014} are plotted in Fig.~\ref{fig2}(e). From the dispersions, $k_{1}=89.9$~rad/$\mu$m, $k_{2}=93.9$~rad/$\mu$m and $\Delta k=4.0$~rad/$\mu$m, or $\phi_k=\Delta k\cdot d=6.0$, which is reasonably close to $2\pi$ considering the uncertainty in $d$ introduced by the finite width $w\approx100$~nm of the wires. Micromagnetic simulations provide further support for coherent phase transfer between the Co magnetizations~\cite{SI}. Figures~\ref{fig2}(f) and~\ref{fig2}(g) show the dynamics of the $y$ component of the wire magnetization as indicated in the insets, which precesses either in phase $\Delta\phi=0$ (Fig.~\ref{fig2}(f)) or with $\Delta\phi=\pi/2$ (Fig.~\ref{fig2}(g)), as a function of $H$ and $f$. 

\begin{figure}[!ht]
\includegraphics[width=86mm]{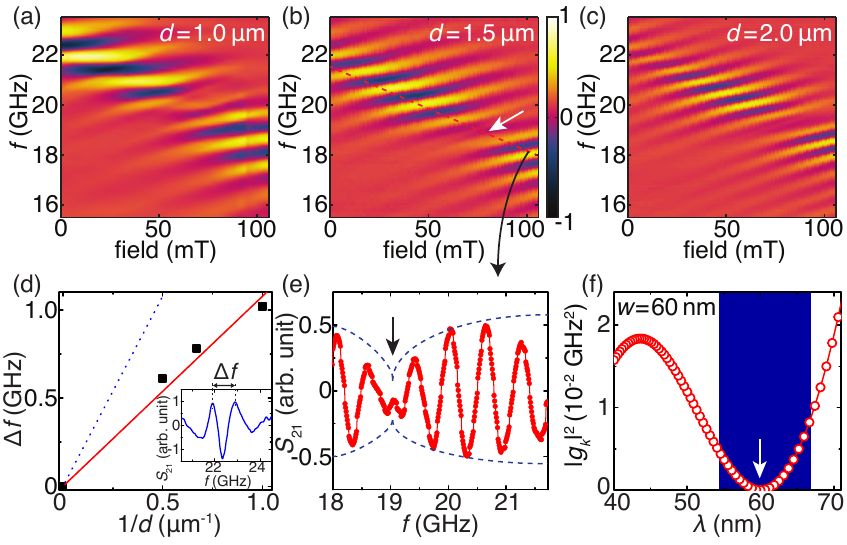}
\centering
\caption{(a-c) Transmission spectra $S_{21}$ measured on the sample with nanowire width $w=60$~nm at different propagation distances, namely $d=1.0$~$\mu$m , 1.5~$\mu$m and 2.0~$\mu$m. The color scales are identical and shown in (b) only. (d) The frequency intervals $\Delta f$ extracted at the Co FMR (the inset is from (a)) are plotted as a function of $1/d$. The red solid line and blue dotted line are calculations with and without taking the additional phase jump $\phi_{\text{i}}$ at the contacts into account. (e) A cut of the spectrum along the FMR (red dashed line) in (b). The blue dashed envelope is a guide to the eye and the arrow indicates the minimum around 19~GHz. (f) Relative coupling strength calculated as a function of wavelength $\lambda$. The blue region shows the range ($\lambda=54.0\sim66.8$~nm) corresponding to (e). }
\label{fig3}
\end{figure}

$S_{21}$ depends on the propagation distance $d$ via the spin-wave propagation phase delay $\phi_{k}=kd$ as illustrated in Figs.~\ref{fig3}(a-c) for a wire width $w=60$ nm~\cite{SI} The spin-wave group velocity $v_{\text{g}}=\frac{d\omega}{dk}=\frac{2\pi\Delta f}{k}=\frac{2\pi}{\Delta\phi_{k}}\Delta f\cdot d$, where $\Delta f$ is the peak-to-peak frequency difference as shown e.g. in the inset of Fig.~\ref{fig3}(d) (a cut of Fig.~\ref{fig3}(a) at zero field) that corresponds to $\Delta\phi_{k}=d\Delta k=2\pi$ and $v_{\text{g}}=\Delta f\cdot d$ or $\Delta f=v_{\text{g}}\left(1/d\right)$~\cite{Vla2008,Neu2010,Yu2014}. Figure~\ref{fig3}(d) shows the observed $\Delta f$ as a function of $1/d$ derived from Fig.~\ref{fig3}(a-c). Using the group velocity for dipole-exchange magnons~\cite{Kal1986,SI} leads to a linear relationship between $\Delta f$ and $1/d$ (blue dotted line in Fig.~\ref{fig3}(d)) that deviates significantly from the experimental data. However, taking $\phi_{\text{i}}=\pi/2$ into account, $\phi_{k}=\Delta\phi-2\phi_{\text{i}}=\pi$ and $\Delta f=\frac{1}{2}v_{\text{g}}\left(1/d\right)$. This leads to the red solid line in Fig.~\ref{fig3}(d) that agrees nicely with experimental data. 

$S_{21}$  for $w=60$~nm shows a clear beating pattern for all $d$ that leads to a vanishing oscillation around 19~GHz (white arrow in Fig.~\ref{fig3}(b)) that do not show up for $w=100$~nm sample (Fig.~\ref{fig2}(b)). Indeed, we argue that the interlayer coupling strength $g_{k}$ should depend on the wire width~\cite{SI}. The Hamiltonian for a single wire close to the magnetic film can be written as

\begin{widetext}
\begin{equation}\label{Hamiltonian1}
	\hat{\mathcal{H}}_{\text{Co$|$YIG}}/\hbar=\omega^{\text{Co}}\hat{m}^{\dagger}\hat{m}+\sum_{k}\omega_{k}^{\text{YIG}}\hat{p}_{k}^{\dagger}\hat{p}_{k}+\sum_{k}\left(g_{k}\hat{m}^{\dagger}\hat{p}_{k}+g_{k}^{*}\hat{m}\hat{p}_{k}^{\dagger}\right)
\end{equation}
\end{widetext}
where $\hat{m}$ and $\hat{p}_{k}$ are bosonic operators associated, respectively, with the Kittel mode in Co and spin waves in YIG with wavenumber $k$. $\omega^{\text{Co}}$ is the Co FMR frequency, $\omega_{k}^{\text{YIG}}$ the YIG spin-wave frequency dispersion~\cite{Kal1986,SI}. The interlayer coupling matrix elements $g_{k}$ can be caused by the interface exchange or magnetodipolar interaction~\cite{Weiler2018,Chen2018,Qin2018,Li2020}. The indirect Co-Co coupling strength as sketched in Fig.~\ref{fig2}(a) scales with $|g_{k}|^{2}\propto|\int_{0}^{w}e^{iky}dy|^2=|\frac{\text{sin}\left(kw/2\right)}{k/2}|^2$, which is plotted as a function of wavelength $\lambda=2\pi/k$ in Fig.~\ref{fig3}(f)~\cite{SI}. The blue shaded interval $\lambda=54.0\sim66.8$~nm corresponds to the spectrum plotted in Fig.\ref{fig3}(e)~\cite{SI,Kal1986}. In this range, the relative coupling strength shows a minimum around 60~nm (Fig.~\ref{fig3}(e)), which explains the suppressed amplitude around 19~GHz (Figs.~\ref{fig3}(a-c). In the $w=100$~nm sample data in Fig.~\ref{fig2}(d), we also observe some amplitude modulation~\cite{SI}. The experimental results shown in Fig.~\ref{fig3}(b) may be directly compared with the calculated transmission based on the theoretical analysis~\cite{Gar1985,Cle2010,SI}. 
The coherent excitation of Co2 from Co1 is driven by a magnon spin torque~\cite{Slon2010,Han2019,Wang2019} mediated by exchange spin waves in YIG. The torque transferred at Co1$|$YIG and YIG$|$Co2 may be induced by either interfacial exchange or magnetodipolar coupling. In view of an interfacial exchange coupling strength measured to be 0.2~GHz in planar Co$|$YIG films~\cite{Weiler2018}, the exchange coupling strength at the interface between a single Co nanowire and the YIG film is merely 20~kHz, that is negligible in comparison with the magnetodipolar coupling calculated and shown in Fig.~\ref{fig3}(f). It has been theoretically studied that the interlayer exchange (magnetodipolar) coupling is larger at the parallel (antiparallel) configuration~\cite{TYu2019}. The experimental observation that strong transmission between Co1 and Co2 occurs at the antiparallel configuration (Fig.~\ref{fig1}(c)) indicates that the magnetodipolar interaction is responsible for the coupling at Co1$|$YIG and YIG$|$Co2. The dominance of magnetodipolar coupling and the absence of interlayer exchange give rise to the spin-wave chirality and eventually lead to the non-Hermitian coupling of two Co wires. The coupling between two Co wires can be described by an effective Hamiltonian as
\begin{equation}\label{Hamiltonian2}
\begin{split}
	\hat{\mathcal{H}}_{\text{Co$|$Co}}/\hbar =&~\omega_{1}^{\text{Co}}\hat{m}_{1}^{\dagger}\hat{m}_{1}+\omega_{2}^{\text{Co}}\hat{m}_{2}^{\dagger}\hat{m}_{2}+i\Gamma_{21}\hat{m}_{1}^{\dagger}\hat{m}_{2}\\
	&+i\Gamma_{12}\hat{m}_{1}\hat{m}_{2}^{\dagger}\\
	 \Gamma_{21} =&~\int \frac{dk}{2\pi} \frac{g_{+k,1}g_{+k,2}^{*}}{i\left(\omega-\omega_{k}\right)-\kappa_{\text{YIG}}}\\
	 \Gamma_{12} =&~\int \frac{dk}{2\pi} \frac{g_{-k,1}^{*}g_{-k,2}}{i\left(\omega-\omega_{k}\right)-\kappa_{\text{YIG}}}
\end{split}
\end{equation}
where $\Gamma_{21}$ ($\Gamma_{12}$) is the coupling from Co1~(2) to Co2~(1), $\kappa_{\text{YIG}}$ stands for the FMR linewidth of the YIG film, $g_{+k,1}$ $g_{+k,2}$ the coupling between Co1 (2) and propagating spin waves in YIG with positive wavenumber $+k$, $g_{-k,1}$ ($g_{-k,2}$) the coupling strength between Co1 (2) and spin waves with $-k$. Here the spin-wave coupling is chiral, $g_{-k,1}=0$ and $g_{-k,2}=0$, as illustrated in Fig.~\ref{fig2}(a)~\cite{Chen2019,TYu2019} and $\Gamma_{12}\approx0$ and hence Eq.~\ref{Hamiltonian2} is non-Hermitian. 

We have demonstrated on-chip non-Hermitian coupling of two distant nanoscale magnets at room temperature. The two Co wires are found to be phase-related by propagating exchange spin waves in the YIG film. Any desired phase shift $\Delta\phi$ from 0 to 2$\pi$ can be realized by tuning frequencies and applied magnetic fields in the vicinity of the Co FMR. The magnon transmission varies sensitively with propagation distance and the width of the nanowire, providing additional evidence for the phase coherent coupling. We theoretically describe the non-Hermitian coupling between two Co wires via unidirectional propagating spin waves in the YIG film. The chiral coupling of nanomagnets demonstrated in this work is attractive for neuromorphic computing~\cite{Tor2017}, since it mimics the innately unidirectional information transmission between neurons via synapses~\cite{Kandel2000}.\\

The authors thank L. Flacke, M. Althammer, M. Weiler, K. Schultheiss, H. Schultheiss, M. Madami, G. Gianluca and C.M. Hu for their helpful discussions. Funding: We wish to acknowledge the support by NSF China under Grants No. 11674020 and No. U1801661, the 111 talent program B16001, the National Key Research and Development Program of China Grants No. 2016YFA0300802 and No. 2017YFA0206200. G.B. was supported by the Netherlands Organization for Scientific Research (NWO) and Japan Society for the Promotion of Science Kakenhi Grants-in-Aid for Scientific Research (Grant No. 19H006450). K.X. thanks the National Key Research and Development Program of China (Grants No. 2017YFA0303304 and No. 2018YFB0407601) and the National Natural Science Foundation of China (Grants No. 61774017 and No. 11734004). K.S. was supported by the Fundamental Research Funds for the Central Universities (Grant No. 2018EYT02). C.L. and M.Z.W. were supported by the US National Science Foundation (Grant No. EFMA-1641989).

\providecommand{\noopsort}[1]{}\providecommand{\singleletter}[1]{#1}%

\end{document}